\documentclass[aps, prx,noshowpacs,reprint,superscriptaddress]{revtex4-1}
\usepackage{amsmath}
\usepackage{pifont}
\usepackage{amssymb}
\usepackage{tipa}
\usepackage{amsfonts}
\usepackage{mathrsfs}
\usepackage{graphicx}
\usepackage{dcolumn}
\usepackage{bm}
\usepackage{color}
\usepackage{booktabs}
\usepackage[colorlinks,linkcolor=red,citecolor=blue]{hyperref}
\usepackage{threeparttable}
\usepackage{multirow}
\usepackage{epstopdf}

\begin{document}

\title{\textcolor{blue}{Publisher link: \href{https://doi.org/10.1103/PhysRevApplied.13.054030}{Phys. Rev. Appl. 13, 054030 (2020)}} \\Electrical Contact between an Ultrathin Topological Dirac Semimetal and a\\ Two-Dimensional Material}

\author{Liemao Cao}
\affiliation{Science, Mathematics and Technology (SMT), Singapore University of Technology and Design (SUTD), 8 Somapah Road, Singapore 487372. }
\affiliation{College of Physics and Electronic Engineering, Hengyang Normal University, Hengyang 421002, China}

\author{Guanghui Zhou}
\affiliation{Department of Physics, Key Laboratory for Low-Dimensional Structures and Quantum Manipulation (Ministry of Education), and Synergetic Innovation Center for Quantum Effects and Applications of Hunan, Hunan Normal University, Changsha 410081, China.}

\author{Qingyun Wu}
\affiliation{Science, Mathematics and Technology (SMT), Singapore University of Technology and Design (SUTD), 8 Somapah Road, Singapore 487372. }

\author{Shengyuan A. Yang}
\affiliation{Science, Mathematics and Technology (SMT), Singapore University of Technology and Design (SUTD), 8 Somapah Road, Singapore 487372. }

\author{Hui Ying Yang}
\affiliation{Engineering Product Development (EPD), Singapore University of Technology and Design (SUTD), 8 Somapah Road, Singapore 487372. }

\author{Yee Sin Ang}
\email{yeesin$_$ang@sutd.edu.sg}
\affiliation{Science, Mathematics and Technology (SMT), Singapore University of Technology and Design (SUTD), 8 Somapah Road, Singapore 487372. }

\author{L. K. Ang}
\email{ricky$_$ang@sutd.edu.sg}
\affiliation{Science, Mathematics and Technology (SMT), Singapore University of Technology and Design (SUTD), 8 Somapah Road, Singapore 487372. }

\begin{abstract}
	
	Ultrathin films of topological Dirac semimetal, Na$_3$Bi, has recently been revealed as an unusual electronic materials with field-tunable topological phases.
	Here we investigate the electronic and transport properties of ultrathin Na$_3$Bi as an electrical contact to two-dimensional (2D) metal, i.e. graphene, and 2D semiconductor, i.e. MoS$_2$ and WS$_2$ monolayers.
	Using combined first-principle density functional theory and nonequilibrium Green's function simulation, we show that the electrical coupling between Na$_3$Bi bilayer thin film and graphene results in a notable interlayer charge transfer, thus inducing sizable $n$-type doping in the Na$_3$Bi/graphene heterostructures.
	In the case of MoS$_2$ and WS$_2$ monolayers, the lateral Schottky transport barrier is significantly lower than many commonly studied bulk metals, thus unraveling Na$_3$Bi bilayer as a high-efficiency electrical contact material for 2D semiconductors.
	These findings opens up an avenue of utilizing topological semimetal thin film as electrical contact to 2D materials, and further expands the family of 2D heterostructure devices into the realm of topological materials.

\end{abstract}

\maketitle

\section{Introduction}

Two-dimensional (2D) layered materials and topological semimetals represent two of the most active research fields of current condensed matter physics, material science and applied device engineering.
The ever-expanding family of 2D materials, such as graphene \cite{geim}, transition metal dichalcogenides (TMDCs) \cite{manzeli}, black phosphorus \cite{xia}, group VI elemental Xenes \cite{molle}, MXenes \cite{naguib}, and the ferromagnetic Cr$_2$Ge$_2$Te$_6$ and Fe$_3$GeTe$_2$ \cite{cheng,deng}, has been widely regarded as a key material class of key importance in electronic \cite{fiori}, photonic \cite{xia2}, optoelectronic \cite{QH_wang}, energetic \cite{pomerantseva}, spintronic \cite{roche} and valleytronic \cite{schaibley, ANG_valley} devices with strong potential in revolutionizing the next-generation solid-state technology.
Electrically contacting 2D materials with metal, however, remains one of the major bottlenecks towards the realization of high-performance industrial-grade electron devices. Although having Ohmic contact is more desirable in reducing power dissipation \cite{Allain1,Cui1,Popov}, most 2D semiconductors are inevitably plagued by the formation of Schottky contacts with inherently large contact resistance when contacted by a bulk metal \cite{Kang,angsin,angsin2,wuqinyun}.
Achieving high-quality electrical contact to 2D materials remains an ongoing quest critically important for the development of industrial-grade 2D-material-based devices \cite{roadmap, ferrari}.

\begin{figure*}[ht]
	\includegraphics[width=6.5 in]{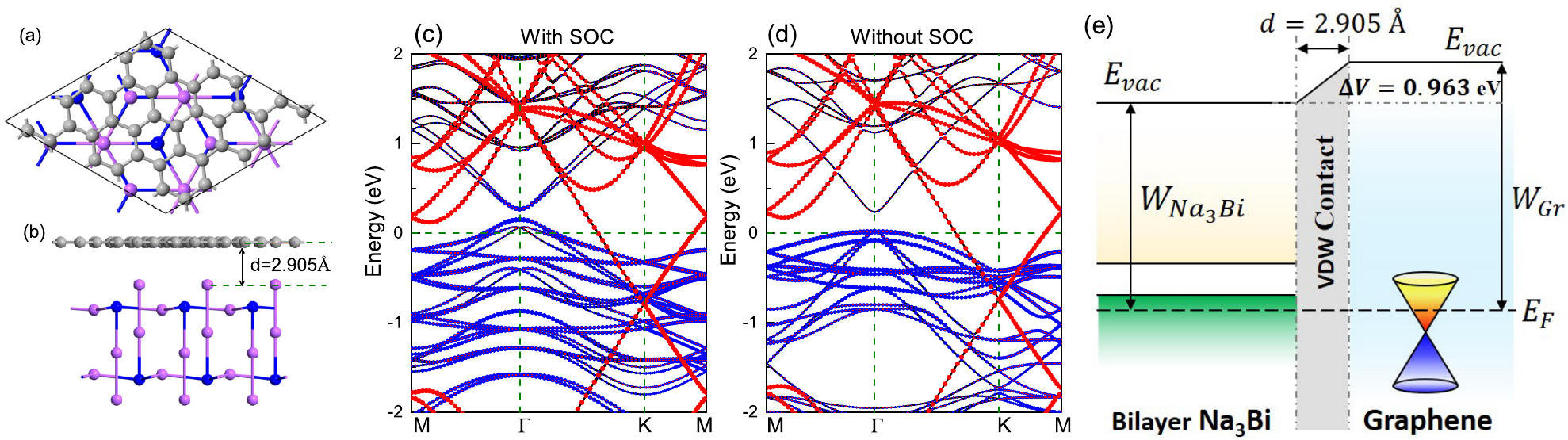}
	\caption{(a) Top view and (b) side view of bilayer-Na$_3$Bi/graphene heterostructure. (c) and (d) Projected band structures with SOC and without SOC, respectively. The blue (red) dot projection originates from bilayer-Na$_3$Bi (graphene). (e) Band alignments of the Graphene/Na$_3$Bi heterostructure. $\Delta$V represents the potential change generated by the interaction between graphene and bilayer-Na$_3$Bi. The work function of Na$_3$Bi and graphene are denoted as $W_{Na_3Bi}$ and $W_{Gr}$, respectively.  }
\end{figure*}

On the other hand, three-dimensional (3D) topological semimetals \cite{Liu1,Neupane,Borisenko} which host exotic topologically protected band crossing in the bulk and topological surface states represents another emerging condensed matter system. For example, 3D topological \emph{Dirac semimetal}, such as Cd$_3$As$_2$ and Na$_3$Bi \cite{Wang,Wang2}, which hosts a pair of 3D massless Dirac quasiparticles have attracted considerable interest in the experimental and theoretical communities \cite{Yang2,Zhang1,Xiong}.
The unusual physical properties of topological semimetals has led to myriads of unusual physical properties, such as the chiral anomaly \cite{Xiong}, linear quantum magnetoresistance \cite{Abrikosov,Zhang2}, oscillating quantum spin Hall effect \cite{Liu2}, exceptional charge carrier mobility exceeding 6000 cm$^2$V$^{-1}$s$^{-1}$ \cite{Hellerstedt}, logarithmically diverging giant diamagnetism at the Dirac nodal point \cite{Goswami,Koshino}, gate-tunable surface states \cite{xiao}, and the strong optical nonlinearity \cite{ANG_NOR}.
Intriguingly, Na$_3$Bi becomes a topological insulators with a bandgap of $\sim300$ meV in the ultrathin limit of a few atomic layers \cite{Collins} in which the band structure can be tuned between topological insulator and band insulator via an external electric field \cite{Pan}. The experimental demonstration of field-effect tunable topological phase transition in Na$_3$Bi bilayer \cite{Collins} thus opens up possibilities of \emph{topological} nano-device in which the charge conduction is controlled by the field-effect modulation of the \emph{band topology} of the transport channel.

Although both 2D materials and 3D topological semimetals have been extensively studied, the union of topological Dirac semimetals and 2D material in terms of contact engineering remains largely unknown thus far.
In this work, we explore the integration of topological semimetal ultrathin film into the design of 2D-material-based electronic devices.
We propose a previously unexplored concept of utilizing ultrathin films of topological semimetals as aelectrical contact to 2D materials.
Using a field-effect transistor setup, we show that the Schottky barrier height (SBH) formed at the bilayer-Na$_3$Bi/2D-semiconductor is markedly lower than that formed by many bulk metals (see Fig. 6 below), thus unraveling the potential of Na$_3$Bi ultrathin film as an efficient electrode to 2D semiconductors.
Furthermore, unlike many bulk metals that \emph{metalizes} the 2D semiconductor during contact formation, the electronic band structures of MoS$_2$ and WS$_2$ remain intact when contacted by Na$_3$Bi, thus offering a useful platform for the optoelectrical studies of optics, valley and excitonic physics in 2D TMDCs \cite{eginligil, j_lee}.
Importantly, ultrathin Na$_3$Bi undergoes metal-to-insulator transition, and a topological phase transition from trivial band insulator to topological insulator with conducting edge states, when it is subjected to an external gate-voltage \cite{Collins,Pan}.
Such gate-tunable electronic properties may be further harnessed as a functional control for designing electronic devices not found in normal metal-semiconductor contact.
Our findings shall open up an avenue towards sub-$10$ nm device technology via the union of 2D materials and the ultrathin films of 3D topological materials, and further expands the family of 2D-material-based heterostructure devices into the realm of quantum materials with nontrivial topological phases.

\section{Computational Methods}

We perform a first-principle density functional theory (DFT) simulation as implemented in the VASP code\cite{Kresse, Kresse1} to study the contact between ultrathin bilayer Na$_3$Bi with: (i) semimetallic graphene; and (ii) semiconducting MoS$_2$ and WS$_2$ monolayers.
The geometric optimization and electronic properties of the heterostructures were calculated via DFT simulation as implemented in the VASP code \cite{Kresse,Kresse1}.
The calculation uses the projector augmented-wave (PAW) pseudopotential \cite{Kresse2}. During the structure relaxations, the spin-orbit coupling (SOC) is not included.
The SOC is included in the electronic structure calculations.
The cutoff energy set to 500 eV. The vacuum layer of 30 \AA $ $ in the vertical direction was chosen to avoid the surfaces interaction between the top and bottom.
Monkhorst-Pack $k$-point sampling is set to be $11\times11\times1$. The method of Grimme (DFT-D3) for vdW interaction functional is adopted in the calculations \cite{Grimme}, and the dipole correction was also added. The residual force was converged to smaller than 0.01 eV/\AA, and the convergence criterion of total energy is set to be 10$^{-6}$ eV. In addition, the transport properties for the two-probe systems are calculated using the Atomistix Toolkit (ATK) 2018 package based on DFT combined with the nonequilibrum Green's functions (NEGF) \cite{Soler,Brandbyge,Taylor,Taylor1}.
SOC is also included in the transport simulation.

\section{Results and Discussions}

\subsection{Bilayer-Na$_3$Bi/graphene heterostructure}

Figures \textcolor{blue}{1(a)} and \textcolor{blue}{1(b)} illustrate the graphene/bilayer-Na$_3$Bi heterostructure in top and side views, respectively.
The optimized lattice parameters of bilayer Na$_3$Bi are $a=b=5.448 $ ${}$\AA, which agrees with the previously reported values \cite{Wang}, and that of the graphene is 2.46 ${}$\AA.
We consider a supercell with (2$\times$2) and ($\sqrt{19}\times\sqrt{19}$) periodicity for Na$_3$Bi bilayer and graphene, respectively.
The lattice mismatch is 0.79 \%, and we stretch the graphene system while keeping the Na$_3$Bi lattice fixed so to preserve the electronic properties of Na$_3$Bi.
The energetically most stable configuration, depicted in Fig.\textcolor{blue}{1(a)}, is obtained by optimizing the lateral displacements of the graphene along the $x$ and $y$ directions with the vertical distance between the bilayer-Na$_3$Bi and graphene (see Fig. \textcolor{blue}{S1} of Supplemental Material \cite{SM}).
The distance of the van der Waals (VDW) gap is 2.905 \AA$ $ at equilibrium.

The projected electronic band structure with and without SOC for the graphene/Na$_3$Bi heterostructure shown in Fig. \textcolor{blue}{1(c)} and \textcolor{blue}{1(d)}, respectively.
The electronic band structures of both Na$_3$Bi and graphene are well-preserved upon forming the heterostructure while the band gap of the bilayer Na$_3$Bi is modified from 300 meV to 121 meV when SOC is included.
Here, graphene is $n$-doped because of the larger work function of graphene that leads to the transfer of electrons from Na$_3$Bi to graphene.
Such transfer also causes some of the valence band of the Na$_3$Bi to cross the Fermi level (see Fig. \textcolor{blue}{1(e)} for the band alignments of bilayer-Na$_3$Bi/graphene contact).
Similar $n$-type doping has also been identified in graphene/Cd$_3$As$_2$ heterostructure \cite{Wu}, thus suggesting the $n$-doping as a common feature in graphene/ultrathin-Dirac-semimetal heterostructures.
In contrast, dominantly $p$-type doping have been reported in graphene contacted by bulk metals \cite{Giovannetti,Khomyakov,Diaz}.

\begin{figure}[t]
	\includegraphics[width=3.5 in]{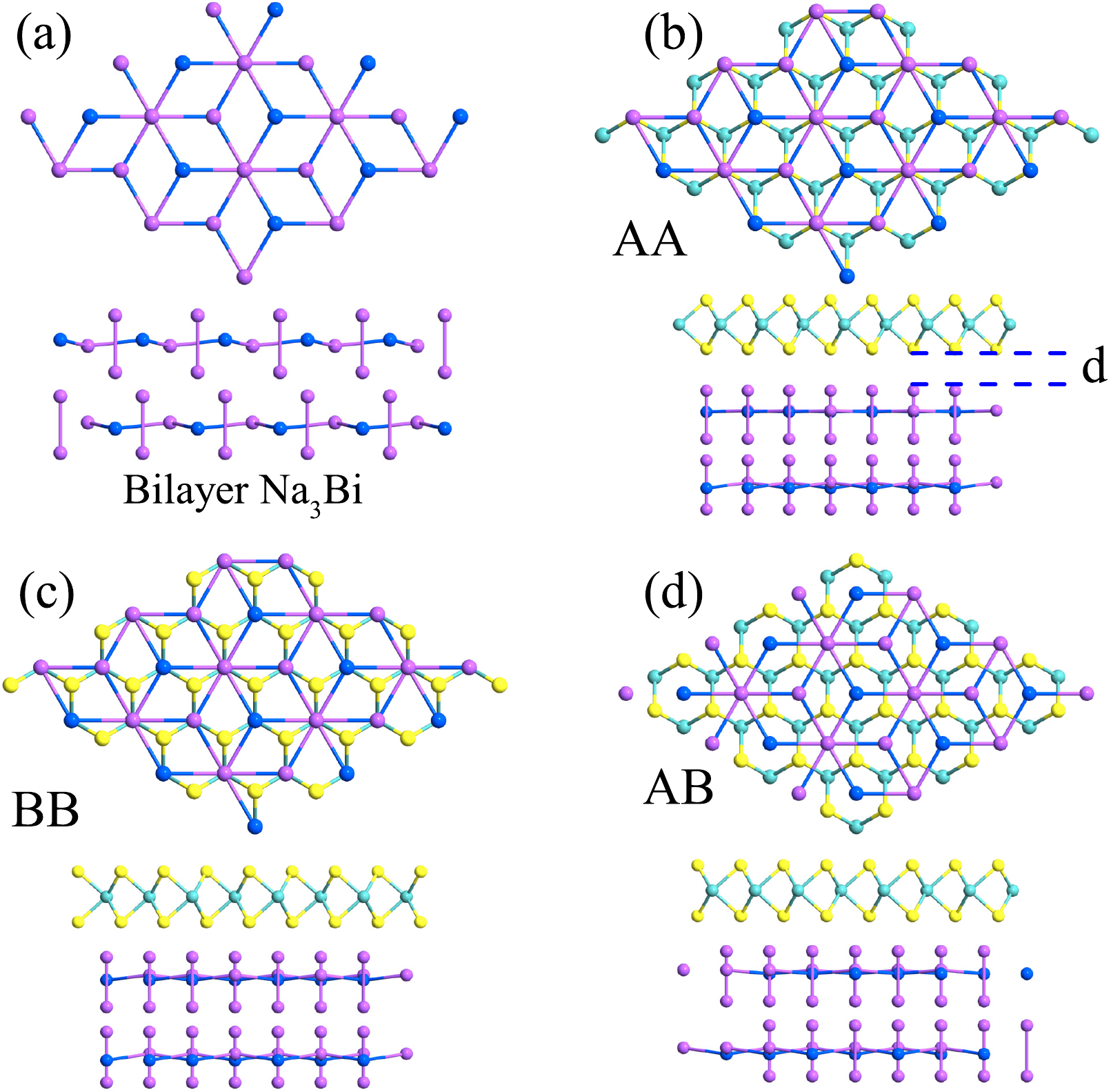}
	\caption{(a) Optimized structural geometry of bilayer-Na$_3$Bi in top (top panel) and side view (bottom panel). (b)-(d) Same as (a) but with different stacking configurations for Na$_3$Bi/MoS$_2$ and Na$_3$Bi/WS$_2$. \emph{d} is the interlayer distance.}
\end{figure}

\subsection{Bilayer-Na$_3$Bi/MoS$_2$ and bilayer-Na$_3$Bi/WS$_2$ heterostructures}

\begin{figure}[t]
	\includegraphics[width=3.2 in]{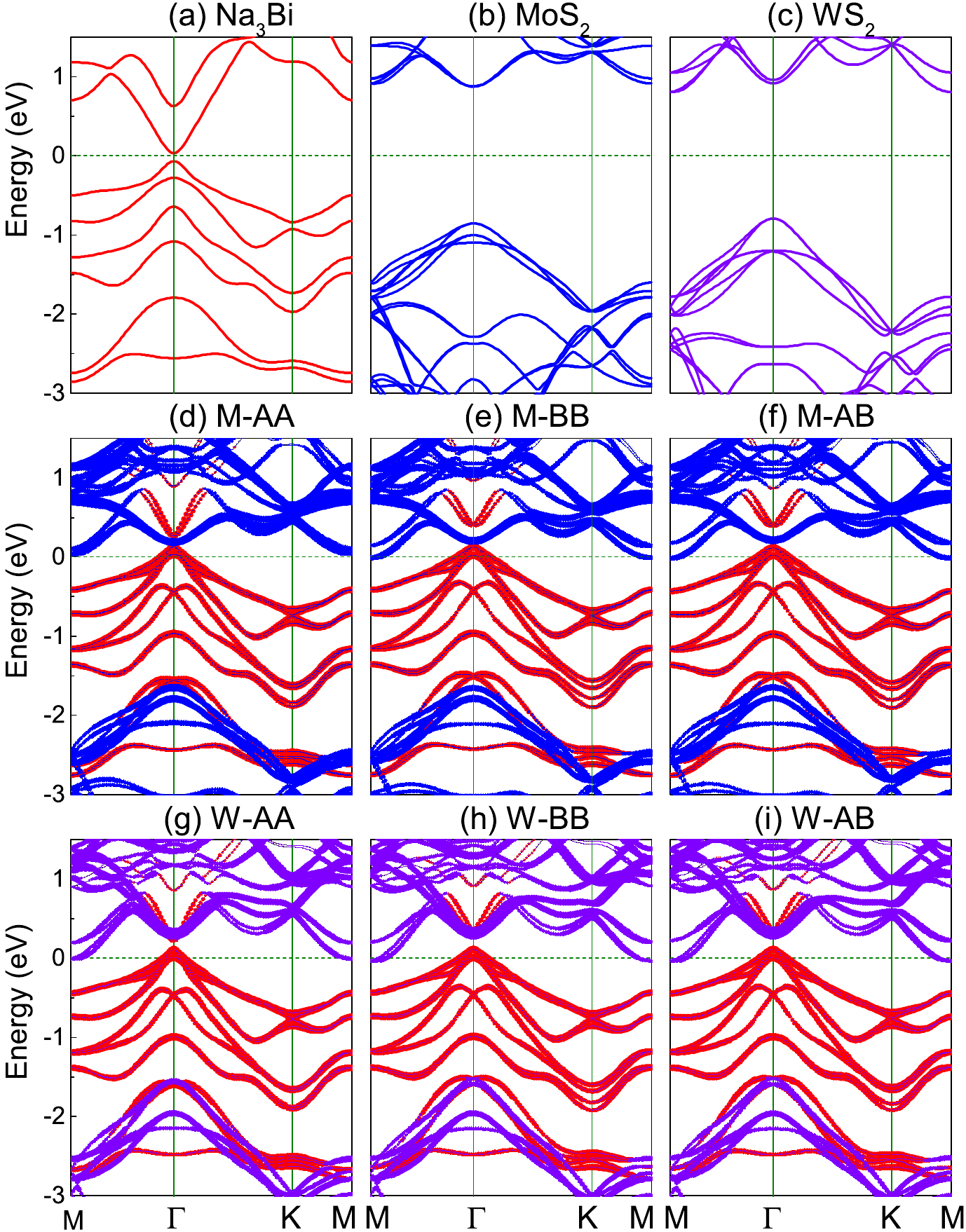}
	\caption{The band structural of bilayer Na$_3$Bi (a), $\sqrt{3} \times \sqrt{3}$ MoS$_2$ (b), and $\sqrt{3} \times \sqrt{3}$ WS$_2$ (c), respectively. (d)-(f) Projected band structures of the heterostructure with SOC for Na$_3$Bi/MoS$_2$ vdWHs by different stack. (g)-(i) Projected band structures of the heterostructure with SOC for Na$_3$Bi/WS$_2$ vdWHs by different stack.}
\end{figure}

\begin{table}
	\begin{threeparttable}
		\caption{Calculated parameters of the Na$_3$Bi/2D-semiconductor heterostructures. $d$ is the interlayer distance. $d_a$ is the minimum distance between Na and S atoms. $E_b$ is the binding energy. $E_{g(N)}$ and $E_{g(S)}$ are the band gap for Na$_3$Bi and MoS$_2$/WS$_2$ after contact, respectively. $D$ is the dipole moment, $W$ represent the work function of the system.}
		\begin{tabular}{ccccccccc}
			\toprule\toprule
			&&
			\multicolumn{3}{c}{Na$_3$Bi/MoS$_2$}&&\multicolumn{3}{c}{ Na$_3$Bi/WS$_2$}\\
			\cmidrule(lr){2-6} \cmidrule(lr){7-9}
			Stacking&&AA&BB&AB&&AA&BB&AB\\
			\midrule
			$d $(\AA)&&2.665&2.186&2.198&&2.696&2.257&2.269\cr
			$d_a $(\AA)&&2.665&2.84&2.85&&2.696&2.90&2.90\cr
			$E_b$ (eV)&&$-$3.667&$-$3.831&$-$3.827&&$-$3.543&$-$3.673&$-$3.668\cr
			$E_{g(N)}$ (eV)&&0.131&0.274&0.278&&0.116&0.279&0.249\cr
			$E_{g(S)}$ (eV)&&1.701&1.653&1.709&&1.558&1.547&1.522\cr
			$D$ (Debye)&&1.063&1.091&1.099&&0.873&0.929&0.919\cr
			$W$ (eV)&&4.236&4.265&4.310&&3.975&4.005&4.001\cr
			\bottomrule\bottomrule
		\end{tabular}
	\end{threeparttable}
\end{table}

We next study the contact between bilayer Na$_3$Bi and the monolayers of MoS$_2$ and WS$_2$ (see Fig. \textcolor{blue}{2}).
The Na$_3$Bi/MoS$_2$ and Na$_3$Bi/WS$_2$ heterostructures are constructed via stacking along the $z$-direction.
For MoS$_2$ and WS$_2$, the lattice parameters are $a=b=3.182$ \AA${}$ and 3.180 ${}$\AA, respectively.
The ($\sqrt{3} \times \sqrt{3}$) unit cells of MoS$_2$ and WS$_2$ were adjusted to match the ($1\times1$) unit cell of bilayer Na$_3$Bi.
The lattice mismatches are 0.77\% and 0.73\%, respectively.
In Figs. \textcolor{blue}{2(b)-2(d)}, we show the three possible stacking modes, namely the AA, BB, and AB stacking configurations of Na$_3$Bi/MoS$_2$ as a representative example as the WS$_2$-based contact exhibits the same stacking structures.
Here, all S (yellow) [Mo (dark cyan)] atoms are above the Na (purple) and Bi (blue) atoms in AA (BB) cases and all atoms of Na$_3$Bi lie directly over the center of a hexagon in the upper MoS$_2$ (or WS$_2$) sheet in the AB cases.
The total energy of the Na$_3$Bi/MoS$_2$ and Na$_3$Bi/WS$_2$ heterostructures at different values of interlayer distances is shown in Fig. \textcolor{blue}{S2} \cite{SM}.
After structural relaxation, we find that the different stacking modes exhibit various interlayer distances (see Table 1).
As the interlayer distances are close to the sum of the covalent radii of Na and S atoms, covalent bonds are expected to form between the two layers.
To examine the energetic stability of Na$_3$Bi/MoS$_2$ and Na$_3$Bi/WS$_2$ electrical contacts, the binding energy $E_b$ is calculated as $E_b=E_{Na_3Bi/MoS_2 (WS_2)}-E_{Na_3Bi}-E_{MoS_2 (WS_2)}$, where $E_{Na_3Bi/MoS_2}$ represent the total energies of the heterostructure, $E_{Na_3Bi}$ and $E_{MoS_2 (WS_2)}$ represent the energy of the isolated Na$_3$Bi and MoS$_2$ (WS$_2$), respectively.
The calculated binding energies have negative values, thus indicating that the structures considered in our simulations are energetically stable.
The calculated heterostructure parameters are summarized in Table \textcolor{blue}{I}.
[see Supplemental Material for the calculated work functions, spatial electrostatic potential distributions of BB stack Na$_3$Bi-MoS$_2$/WS$_2$ \cite{SM}]

The band structures of an isolated bilayer Na$_3$Bi and $\sqrt{3} \times \sqrt{3}$ MoS$_2$ (WS$_2$) monolayer are shown in Figs. \textcolor{blue}{3(a)-(c)}.
The bilayer Na$_3$Bi exhibits a direct band gap of 0.1 eV at the $\Gamma$ point which is consistent with previous results \cite{Collins}.
The band gap is 1.73 eV for MoS$_2$, and 1.6 eV for WS$_2$, which are slightly less than those reported in previous studies due to the presence of strain in the heterostructures considered here \cite{Chen}.
The projected band structures of the electrical contacts are shown in Figs. \textcolor{blue}{3(d)-(i)} (see Fig. \textcolor{blue}{S3} for the projected band structures without SOC in \cite{SM}).
Here, red, blue and violet symbols denote the contributions from the Na$_3$Bi, MoS$_2$ and WS$_2$, respectively.
Compared with the band structures of isolated bilayer Na$_3$Bi, $\sqrt{3} \times \sqrt{3}$ MoS$_2$ and WS$_2$ [Figs. \textcolor{blue}{3(a)-(c)}], the band structures of Na$_3$Bi are not only up-shifted and show metallic behavior, but also exhibit obvious band splitting.
Strong splitting of the valence band around $-0.5$ eV at the $\Gamma$ point is observed, which originates from the interactions between Na$_3$Bi bilayer and 2D TMDCs when forming the hereostructure.
For both MoS$_2$ and WS$_2$, their valance bands are energetically down-shifted.

Interestingly, the conduction band minimum (CBM) is shifted from the $\Gamma$ point to the $M$ point, and crosses the Fermi level, thus indicating the formation of an ohmic contact across the VDW gap [see Figs. \textcolor{blue}{3(g)-(i)}].
The ohmic contacts persists even in the case of 10-layer-Na$_3$Bi ($\sim 5$ nm) contacts to MoS$_2$ and WS$_2$ (see Fig. \textcolor{blue}{S4} in \cite{SM}).
Such ohmic vertical interface is in stark contrast to the case of 2D TMDCs contacted by Au and Ag where a Schottky barrier is formed across the vertical VDW gap \cite{Tang,Gong,panyuanyuan}, which impedes the efficiency of charge injection.
For other bulk metal electrodes such as Sc, Ti and Pt, the band structures of 2D TMDCs are metalized in the contact region \cite{Tang,Gong,panyuanyuan}.
This is again in contrast to the case of Na$_3$Bi contacts to 2D TMDCs where the hybridization between the two materials is weak.
The band structures of 2D TMDCs are retained without the formation of Schottky barrier across the vertical contact interface.
Such Ohmic nature of the Na$_3$Bi-contacted MoS$_2$ and WS$_2$ suggests that electrons can be more efficiently injected through the vertical interfaces -- a favorable characteristic for electronics and optoelectronics applications.

\begin{figure}
	\center
	\includegraphics[bb=40 120 760 476, width=3.5 in]{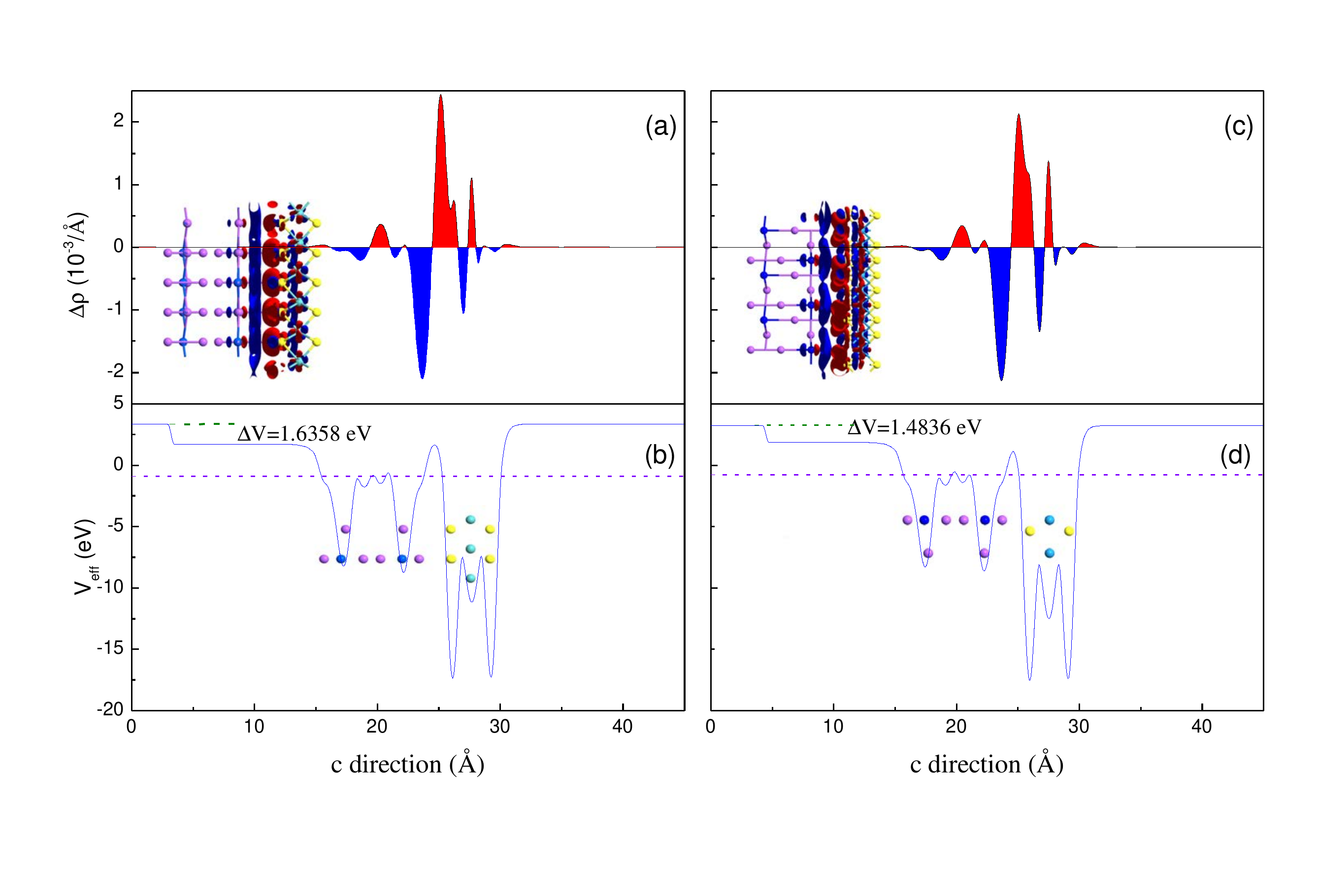}
	\renewcommand{\figurename}{FIG.}
	\caption{Plane-averaged differential charge density $\Delta\rho$ and electrostatic potentials of the Na$_3$Bi with (a, b) MoS$_2$; and (c, d) WS$_2$ heterostructure at the equilibrium along $z$ direction, respectively. The inset shows the side views of the isosurfaces of differential charge density of the heterostructure. The $\Delta$V reflects the difference between the work function on the Na$_3$Bi side and on the 2D semiconductor side.}
\end{figure}

To further understand the detailed nature of the charge transfer at the Na$_3$Bi and MoS$_2$/WS$_2$ interfaces, we calculate the charge difference between the combined heterostructure system and the sum of the isolated Na$_3$Bi and MoS$_2$/WS$_2$ in Figs. \textcolor{blue}{4(a)} and \textcolor{blue}{4(c)}.
The charge density difference is calculated as: $\Delta\rho=\rho_{H}-\rho_{Na_3Bi}-\rho_{MoS_2/WS_2}$, where the $\rho_{H}$, $\rho_{Na_3Bi}$, $\rho_{MoS_2/WS_2}$ are the charge density of the heterostructure, freestanding bilayer Na$_3$Bi, and isolated MoS$_2$ or WS$_2$, respectively.
The blue regions represent electron depletion, while the red region represents the accumulation of electrons in the heterostructures relative to their two isolated components.
At the interfacial region, several charge transfer oscillations are observed.
The main charge depletion is contributed by the first layer of the Na$_3$Bi closest to MoS$_2$ or WS$_2$.
Moreover, some extra charge is found to be accumulating around the Mo and W atoms.
Compared with metal-MoS$_2$/WS$_2$ contacts, the major difference is the charge accumulation located in MoS$_2$ or WS$_2$ interfaces \cite{Gong,Chen}.
In general, both the charge depletion and the charge accumulation constitute the interface charge redistribution behavior, leading to electron wave function polarization and, hence, the formation of interfacial electric dipole.
The presence of such interface dipole directly modifies the interfacial band alignment \cite{Tung}, thus leading to the band structures modification show in Figs. \textcolor{blue}{3(d)-(i)} in the main text.


\begin{figure}[t]
	\includegraphics[bb=170 90 480 520, width=3.2 in]{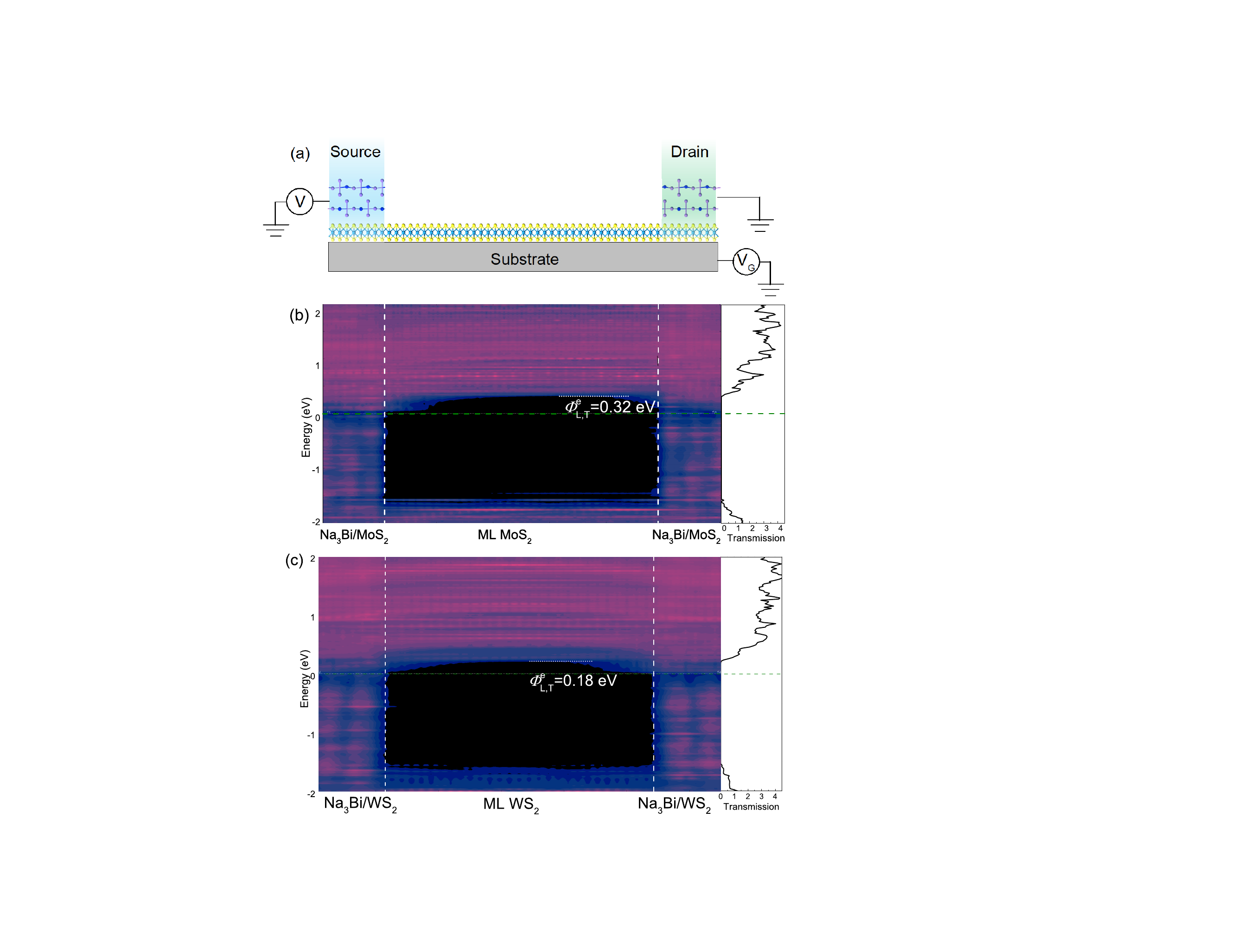}
	\caption{(a) Schematic of a 2D TMDCs field-effect transistor. (b) and (c) LDDOS (left panels) and transmission spectra (right panels) of MoS$_2$ and WS$_2$ transistor with Na$_3$Bi electrodes at zero gate and bias voltages. The green horizontal line represent the Fermi level.}
\end{figure}

\begin{figure}[t]
	\includegraphics[bb=8 10 180 160, width=3.2 in]{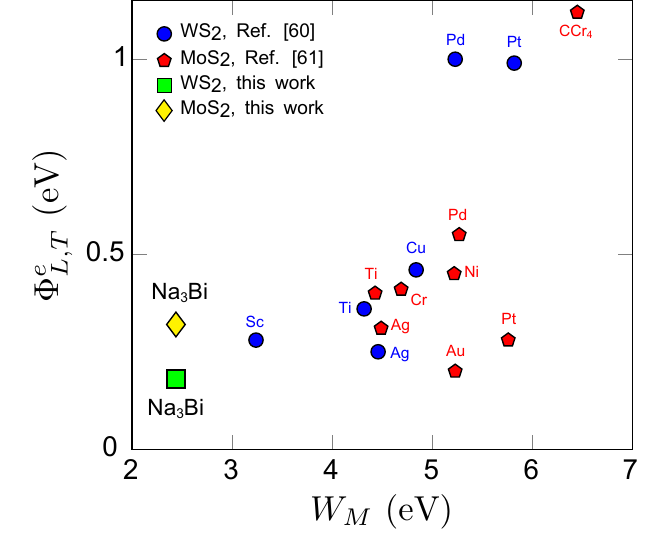}
	\caption{Lateral SBH ($\Phi_{L,T}^e$) as a function of isolated metal work function ($W_M$) for WS$_2$ \cite{Tang} and MoS$_2$ \cite{panyuanyuan}. }
\end{figure}

To illustrate the key strengths of ultrathin Na$_3$Bi as an electrical contact to 2D TMDCs over other bulk metals, we simulate a field-effect transistor device with a 5-nm channel length [see \textcolor{blue}{Fig. 5(a)}].
We employ a nonequilibrium Green's function (NEGF) approach \cite{SM} to extract the \emph{lateral transport Schottky barrier height} laterally extending into the 2D channel by calculating the local device density of states (LDDOS) which reflects the energy-band distribution in real-space along the simulated device \cite{Das,Du}.
The lateral electron SBH ($\Phi^e_{L,T}$) of the MoS$_2$ and  WS$_2$ transistors are estimated as the energy difference between the Fermi level, $E_F$, and the CBM of the 2D channel [labeled in Figs. \textcolor{blue}{5(b)} and \textcolor{blue}{(c)}].
The $\Phi^e_{L,T}$ is calculated as 0.32 eV and 0.18 eV, respectively, for MoS$_2$ and WS$_2$ [Figs. \textcolor{blue}{5(b)} and \textcolor{blue}{(c)}], which are appreciably lower than many other commonly studied bulk metals \cite{Tang,panyuanyuan} because of the low work function of Na$_3$Bi bilayer ($2.44$ eV).
In Fig. \textcolor{blue}{6}, $\Phi_{L,T}^e$ versus isolated metal work function ($W_M$) is plotted using the data for WS$_2$ and MoS$_2$ 5-nm field-effect transistor extracted from Refs. \cite{Tang,panyuanyuan}.
Here $\Phi_{L,T}^e$ of bilayer-Na$_3$Bi is substantially lower when compared to other bulk metals.
Particularly for WS$_2$ contacted by Na$_3$Bi bilayer, the $\Phi_{L,T}^e$ is the lowest compared to other common bulk metal electrodes (i.e. Sc, Ti, Ag, Cu, Pd, and Pt) \cite{panyuanyuan}.
The relatively low $\Phi_{L,T}^e$ thus reveals the potential of Na$_3$Bi bilayer as another candidate electrode material for achieving high-efficiency charge injection into 2D TMDCs.
We further remark that the pronounce Stark effect in Na$_3$Bi ultrathin film allows the band gap to be tuned and closed, and the electronic band structure to be switched between trivial band insulator and topological insulator by an external gate-voltage \cite{Collins, Pan}.
Such gate-tunable metallic-insulator transition and topological phase transitions may offer a physical mechanism for the design of functional devices, not found in the conventional bulk-metal/semiconductor contact, that worth to be explored in future works.

Finally, we remark that Na$_3$Bi is prone to degradation under ambient air conditions.
Nonetheless, recent experiment \cite{Collins} has successfully employed angle-resolved photoemission spectroscopy and scanning tunneling microscopy to study the field-effect tunable band structure in Na$_3$Bi, revealing a major step forward towards the fabrication of Na$_3$Bi device. More recently, Na$_3$Bi ultrathin film passivated by MgF$_2$ or Si capping layers have been demonstrated to be air-stable and the transport properties remain intact after such passivations \cite{cap}. Encapsulated Na$_3$Bi devices thus provide a potential route towards air-stable hybrid Na$_3$Bi/2D-semiconductor devices.

\section{Conclusion}

In summary, we investigated the electronic and transport properties of the ultrathin topological Dirac semimetal Na$_3$Bi as an electrical contact to graphene, MoS$_2$ and WS$_2$ via first-principle calculations.
We show that the Na$_3$Bi/graphene contact leads to a $n$-type doping in graphene, which can be useful for electronics and optoelectronics applications, such as $p$-$n$ junction and photodetector
For Na$_3$Bi/MoS$_2$ and Na$_3$Bi/WS$_2$, the prevalence of ohmic vertical interface and low lateral Schottky barrier heights indicates the potential of Na$_3$Bi as an energy-efficient electrical contact.
The findings reported here could form the harbinger for the exploration of an emerging class of heterostructure electronic devices \cite{liang} that synergies 2D materials and the ever-expanding family of topological semimetals where nodal point, line, link, chain, double helix, hourglass, surface and many other exotic topological phases are continually being unearthed.

\begin{acknowledgments}
	
This project is funded by Singapore MOE Tier 2 Grant (2018-T2-1-007), A$^*$STAR IRG Grant (IRG A1783c0011), and Hunan Provincial Natural Science Foundation of China (Grant No. 2019JJ50016), and science Foundation of Hengyang Normal University of China (No. 18D26), and the National Natural Science Foundation of China (Grant No. 11774085). All the calculations were carried out using the computational resources provided by the National Supercomputing Centre (NSCC) Singapore.

\end{acknowledgments}

\end{document}